\documentclass{elsarticle}

\biboptions{authoryear}

\usepackage{amssymb}
 \usepackage{amsthm}
\usepackage{amsmath}
\usepackage{xcolor}
\usepackage{ogonek}

\usepackage{nameref,hyperref}

\usepackage{array}

\makeatletter
\newcommand{\thickhline}{%
    \noalign {\ifnum 0=`}\fi \hrule height 2pt
    \futurelet \reserved@a \@xhline
}
\newcolumntype{"}{@{\hskip\tabcolsep\vrule width 2pt\hskip\tabcolsep}}
\makeatother

\journal{Finance Research Letters}

\begin{document}

\begin{frontmatter}

\title{Facebook Drives Behavior of Passive Households in Stock Markets
\tnoteref{mytitlenote}}
\tnotetext[mytitlenote]{This is paper has been published in \emph{Finance Research Letters}: \url{https://doi.org/10.1016/j.frl.2018.03.020}}

\author[mymainaddress]{Milla Siikanen\corref{mycorrespondingauthor}}
\cortext[mycorrespondingauthor]{Corresponding author}
\ead{milla.siikanen@tut.fi}
\author[mymainaddress]{K\k{e}stutis Baltakys}
\author[mymainaddress]{Juho Kanniainen}
\author[mysecondaddress,mythirdaddress]{Ravi Vatrapu}
\author[mysecondaddress,mythirdaddress]{Raghava Mukkamala}
\author[mysecondaddress]{Abid Hussain}

\address[mymainaddress]{DARE Business Data Research Group, Laboratory of Industrial and Information Management, Tampere University of Technology, Finland}
\address[mysecondaddress]{Centre for Business Data Analytics, Copenhagen Business School, Denmark}
\address[mythirdaddress]{Westerdals Oslo School of Arts, Communication and Technology, Norway}

\begin{abstract}
Recent studies using data on social media and stock markets have mainly focused on predicting stock returns. Instead of predicting stock price movements, we examine the relation between Facebook data and investors' decision making in stock markets with a unique data on investors' transactions on Nokia. We find  that the decisions to buy versus sell are associated with Facebook data especially for passive households and also for nonprofit organizations. At the same time, it seems that more sophisticated investors---financial and insurance institutions---are behaving independently from Facebook activities. 

\end{abstract}


\end{frontmatter}

\section{Introduction}\label{intro}
Social media sites, such as Facebook and Twitter, create various opportunities for companies to improve their internal and external communications and to collaborate and communicate with their customers, partners, and  other stakeholders, such as investors. Given the importance of social media in external communications, it is not surprising that social media data have been  used recently to predict real-world outcomes \citep[see e.g.][]{asur2010predicting}. In the financial market research, numerous scholars have used Facebook data \citep{karabulut2013can,siganos2014facebook,bukovina15} and data from other social media sites \citep{bollen11,zhang2011predicting,zheludev2014can,chen2014wisdom,nofer2015using,zhang2017celebrities,you2017twitter}.\footnote{See also \citet{bukovina16} for an overview of research related to a link between social media and capital markets} The primary aim of such research has been to predict market-wide stock movements, yet there is scant research on how social media data relate to the behavior of {\em individual investors}, perhaps because of the lack of availability of investor account level data.  

In this paper, we examine the extent to which  investors' trading decisions are driven by Facebook posts and activity. To this end, we use a unique investor-level shareholding registration data set that includes the trading of all Finnish investors over multiple years. In particular, given that an investor trades, we study how Facebook data relate to investors' decisions to increase or decrease their positions. This question is addressed for different investor groups, including financial institutions, nonprofit organizations, and households, and their trades in Nokia stock. As Nokia was one of the most liquid stocks on the Finnish stock market, this unique data has been studied in several articles,\footnote{See for example \citet{westerholm2009uninformed,tumminello2012identification,lillo15,ranganathan2017dynamics}} and here we combine it with social media data. Paper by \citet{lillo15} is the most closely related study to ours. It also investigates the trading behavior of different investor groups with Nokia stock, but with Thomson Reuters news articles---which are not social media data per se. 

Currently, Facebook is clearly the most widely used social media platform, with  2.2 billion monthly active users worldwide \citep{statista}. 
 As of January 2013, social media sites such as Facebook and Twitter are used by about 45\% of S\&P1500 firms to communicate externally formal and informal information about their business \citep{jung2017firms}. Specifically, companies communicate both corporate disclosures and other information via social media \citep{zhou2014social}. \citet{yang2017impacts} show that social media, and mass media in general, influences investor's trading decisions. \citet{snow2017if} argue that less sophisticated  investors potentially benefit most from disclosures communicated via social media, because, on social media platforms, the information is essentially ``pushed'' to them, which makes this information easier to access. In addition, \citet{snow2017if} show that less sophisticated investors process financial information received from social media differently from information received via company's investor relations website.

It is important to remember that typically companies use official exchange-routed company announcements as a primary communication channel \citep[see e.g.][]{jung2017firms}, followed by other channels, including newspapers and social media.\footnote{See \citet{siikanen2017limit,siikanen2017drives}, and references therein, for effects of company announcements in stock markets.} Additionally, communicating information via social media is voluntary, while some company announcement releases are mandatory. Furthermore, \citet{jung2017firms} show that companies disseminate strategically, i.e. companies are less likely to disseminate information in Twitter when the news is bad. In this regard, we wish to determine how the investment decisions of, for example,  less sophisticated and professional investors, among other investor groups, correlate with potentially biased Facebook information. We also note that the relationship between Facebook data and trading can also be related to the attention grabbing behavior of investors, especially households \citep[see][]{barber2007all}.

\section{Data}\label{data}
\subsection{Shareholding Registration Record Data}\label{EC}

To identify the trading of different investor categories, we use shareholding registration record data including all domestic investors from June 7, 2010  to the end of 2016, obtained from Euroclear Ltd.\footnote{\citet{grinblatt00,grinblatt01,tumminello2012identification,lillo15,baltakys2017multilayer} use data sets from the same source, and provide descriptions of the data. However, they use data from before 2009, when all transactions were reported separately with exact trading dates. After moving to Central Counterparty Clearing in late 2009, the Euroclear research data set contains only \emph{aggregated} daily trades \emph{without specifying the actual trading dates}---instead a registration date is reported for each record. Thus, we reverse engineer the trading dates from the registration dates. We use the official T+3 settlement convention for data before and on October 8, 2014 and T+2 afterwards \citep[see][]{euroclear}. Using the derived trading dates, we aggregate transactions on a weekly basis, and this reduced the possible noise of inaccurate trading date derivation.
} Each record in the data contains detailed information about the investor and the change in his/her holdings. During our analysis period, 282,269  distinct Finnish investors traded Nokia stock. We divide them into five groups according to their sector codes: nonfinancial corporations, financial and insurance corporations, general governmental organizations, nonprofit organizations, and households. Household investors are further divided into four investor activity groups. Investor's activity group is defined by the number of days the investor traded during the past eight weeks, including the analyzed week. If the number of active days in the past 8 weeks is equal to 1, the investor is considered inactive; if it is between 2 and 5, the investor is passive; 6--20 means moderate; and 21--40 means active. Notably, this is a dynamic group, as one investor might appear in several groups throughout the analysis period.

For the purposes of our analysis, we calculate the number of investors in each group who changed their holdings during a week and the number of investors who increased their holdings (bought more than sold) during that week. Table \ref{TAB:groupnum} gives the descriptive statistics of the investor groups and their weekly trading in our data sample. We see that financial and governmental institutions are on average most active sector groups, where as households and nonprofit organizations are least active.

\begin{table}[!ht]
\centering
\caption{\textbf{Descriptive statistics on investor groups.} N gives the total number of investors per group. Mean, median and standard deviation (st.Dev) relate to the weekly observations on numbers of investors in each group that changed their net holdings during a week. In Panel B, household investors are categorized into activeness groups on the basis of their trading in the past eight weeks (40 trading days).}
\label{TAB:groupnum}
\footnotesize{
\begin{tabular}{lrrrr}
\hline
\multicolumn{ 5}{c}{{\bf Panel A: Investor categories}} \\
\thickhline
{\bf Sector} &    {\bf N} & {\bf Mean (\%of all)} &  {\bf Median} &  {\bf st.Dev} \\
\hline
 Companies &     12,213 &        271 (2.2\%) & 230 & 166 \\

 Financial &        427 &         28 (6.6\%) & 27 & 9 \\

Governmental &         89 &          7 (7.9\%) & 7 & 4 \\

  Nonprofit &      1,177 &         18 (1.5\%) & 16 & 12 \\

Households &    268,363 &      4,640 (1.7\%) & 3,694 & 3,179 \\
\hline
     Total &    282,269 &   &  &         \\
\thickhline
\multicolumn{ 5}{c}{{\bf Panel B: Activity groups of household investors}} \\
\thickhline
{\bf Activeness; \# of active days} &    {\bf N} &  {\bf Mean (\%of all)} &  {\bf Median} &  {\bf st.Dev} \\
\hline
Active; $(20, 40]$ &      1,228 &         54 (4.4\%)&  51 & 22 \\

Moderate; $(5, 20]$ &     16,019 &        502 (3.1\%)& 450 & 227 \\

Passive; $(1, 5]$ &    120,906 &      1,856 (1.5\%)& 1,402 & 1,422 \\

Inactive; $1$ &    264,942 &      2,228 (0.8\%)& 1,670 & 1,897 \\
\hline
\end{tabular} } 

\end{table}

\subsection{Facebook Data}

We collect daily numbers of posts and related comments, likes, and shares from Nokia's Facebook wall\footnote{\url{https://www.facebook.com/nokia}} between June 2010 and December 2016 using the Social Data Analytics Tool (SODATO) \citep[see][]{hussain14a,hussain14b,hussain14c}. The comments, likes, and shares are always related to a specific post, i.e. the post is the main action. Therefore, we assign the numbers of comments, likes, and shares to the date of the original post---that is, not the date when the actual comment, like, or share was made. In effect, the numbers of comments, likes, and shares quantify the attention the posts released on a particular day received. 

We aggregate the daily Facebook data to weekly by summing the numbers of posts, comments, likes, and shares during a week. We take the week beginning on Saturday and ending on Friday, since trading does not occur on weekends. This way, we relate the Facebook activity on weekends to the week in which they can actually affect investors' trading decisions. In total, our sample comprises of 342 weekly observations for posts, comments, likes, and shares. Table \ref{TAB:avgs} gives descriptive statistics of these time series. We can see that on average, there is more than one post made per day, and calculate that one post got on average 274 comments, 4,379 likes, and 7 shares. 

\begin{table}[!ht]
\centering
\caption{\textbf{Descriptive statistics on Facebook data.} N gives the total number of each Facebook activity in our sample. Mean, median and standard deviation (st.Dev) relate to the weekly observations on numbers of each Facebook activity.}
\label{TAB:avgs} 
\footnotesize{
\begin{tabular}{lrrrr}
 \hline
  \textbf{Activity} &    \textbf{N} &    \textbf{Mean} &\textbf{Median} & \textbf{st.Dev} \\ 
\thickhline
      Post &   2,906 &       8 &     8 &     6 \\ 

   Comment &   797,586 &    2,332 &   1,585 &    2,808 \\ 

      Like &   12,725,171  & 37,208 &     11,977 & 43,500 \\ 

     Share &   919,380 &    2,688 &    461 &   4,525 \\ \hline

\end{tabular} }
\end{table}

\subsection{Company announcement data} The announcement data is collected from NASDAQ OMX Nordic's website.\footnote{\url{http://www.nasdaqomxnordic.com/news/companynews}, see the page also for detailed information.} The data set includes all the announcements that Nokia filed with Nasdaq between June 2010 and December 2016. Altogether, we have 507 company announcements in the sample. We aggregate the announcement data into weekly by summing the number of announcements from Saturday to Friday, i.e. in similar way as the Facebook data. In the regressions, we use a dummy variable to indicate whether there was at least one announcement release during a week. Our sample includes 187  weeks with at least one announcement release (out of total 342 weeks). 

\subsection{Weekly return data} The daily adjusted closing price data used to calculate the returns is collected from NASDAQ OMX Nordic's website.\footnote{\url{http://www.nasdaqomxnordic.com/shares/microsite?Instrument=HEX24311}.} For each week, we calculate the log return as $\text{Ret}_t = \text{ln}\left[P_t / P_{t-1}\right]$, where $P_t$ is the closing price from the last trading day on the week (usually  Friday), and $P_{t-1}$ is the closing price from last trading day on the previous week $t-1$ (usually previous week's Friday). The average weekly return for Nokia during the sample period was --0.16\%.

\section{Framework of the empirical analysis}\label{framework}

Our analysis is based on logistic regressions to explain how Facebook activity relates to an increase versus a decrease in Nokia shares in investors' portfolios.\footnote{Another option would be (instead of restricting the analysis to a binary outcome) to use linear regressions with continuous dependent variable (i.e. how much an investor changed the position). However, in order to use continuous dependent variable, proportional changes in investors' positions would have to be calculated, which, in turn, requires information on investors' holdings. In contrast to changes in holdings, the levels of holdings, however, were not accurately available. The use of ``changes in holdings'' as a non-proportional variable is problematic, because investors are trading by very different amounts of shares. These problems are addressed by using logistic regression.} To identify the groups of investors whose trading behavior is related to Facebook data, we run separate regressions for each investor group with each Facebook variable.

The dependent variable in our regressions is a dummy variable with value 1 if an investor increased his/her holdings in Nokia stock during a given week (bought more than sold) and 0 if the investor decreased the holdings ($D^{\text{increased}}_{t}$). In a given week, only investors whose net position for Nokia changed are included. The explanatory  variable of main interest is the number of posts, comments, likes, or shares depending on the regression (FB). We control for company announcement releases with company announcement dummy (NEWS$_t$), which is 1 if there was an announcement released during week $t$ and 0 otherwise. Additionally, we use the number of investors in the group who increased their holdings during the previous week scaled by the total number of investors who changed their holdings during the previous week. This is depicted as follows: 

\begin{equation*}
\begin{split}
\text{scD}_{t-1}^{\text{increased}} =  \frac{1}{n_{t-1}} \sum_{i=1}^{n_{t-1}} D^{\text{increased}}_{i,t-1}  
\end{split}
\end{equation*} 
where $n_{t-1}$ is the number of investors who changed (increased or decreased) their holdings in Nokia during week $t-1$. We also add control variables for the return on present week (Ret$_t$) and the previous week (Ret$_{t-1}$). Lastly, we include monthly ($M_t$) and yearly ($Y_t$) dummy variables. The monthly dummies control for the potential yearly seasonality in the trading \citep[for example, realizing the losses in December for tax purposes, see e.g.][]{grinblatt01}, and the yearly dummies accommodate the analysis for example to possible changes due to the abandonment of Nokia's mobile business (in 2014, Nokia's mobile business was acquired by Microsoft, changing the focus of the company to a telecommunications infrastructure business). To summarize, the regressions we run are of the following form:
\begin{equation}
\label{EQ:reg}
\begin{split}
g(D^{\text{increased}}_{t}) = \alpha_1 + \alpha_2 \cdot \text{FB}_t + \alpha_{3} \cdot \text{NEWS}_t  + \alpha_{4} \cdot \text{scD}^{\text{increased}}_{t-1} \\ 
+ \alpha_{5} \cdot \text{Ret}_t + \alpha_{6} \cdot \text{Ret}_{t-1} + \sum_{j=1}^{11} \alpha_{j+6} \cdot M_j + \sum_{j=1}^{6} \alpha_{j+17} \cdot Y_j
\end{split}
\end{equation} 
where $g$ is the logit function.

\section{Results}\label{results}

Panel A in Table \ref{TAB:all} shows that for households and nonprofit institutions, all the regression estimates are statistically significant. The results indicate that the decisions of investors in these groups to buy vs. sell have a clear association with the Facebook data. For nonprofit institutions, the economic significance is relatively high: the odds of a nonprofit institution buying rather than selling range from 1.111 to 1.212 when the amount of Facebook activity increases by one standard deviation. For financial institutions, Panel A in Table \ref{TAB:all} shows no association between the buy vs. sell decisions and the Facebook data. The results for companies and governmental institutions are something between those of financial institutions and households and nonprofit institutions, as  half or less of the estimates are statistically significant. 

To take a closer look at the effect of Facebook on the trading of households, Panel B in Table \ref{TAB:all} presents the estimated regression results for individual investors in different activity groups. We observe that, in general, the more active a household is, the weaker is the association between Facebook data and buying/selling behavior. The odds ratios for passive and inactive investors are more modest than those of nonprofit institutions, though for posts they are still relatively high (1.088 and 1.072). For brevity, we do not report the regression estimates for interception and control variables here, but they are available in Online Appendix. In general, most of the estimates for control variables are statistically significant.

\begin{table*}[!ht]
\centering 
\caption{\textbf{Regression estimates: Trading of investor groups and Facebook data.}  The estimates related to Facebook variables of logistic regressions described  in Section \ref{framework} (Equation \ref{EQ:reg}) for all the investor categories. The dependent variable is a dummy variable getting value of 1 if an investor increased his/her holdings during the week, and 0 if the investor decreased the holdings. In addition to the Facebook related variables (for which we report the estimates here), we control for company annoncement releases, number of previous weeks investors changing their position (scales), current and previous weeks returns, and in addition we have monthly and yearly dummies. The regression estimates for control variables (omitted here) are available in Online Appendix. In Panel B, household investors are categorized into activeness groups on the basis of their trading in the past eight weeks (40 trading days). Number of observations (weeks in the analysis) is 341 for all the other regressions, except 332 for group governmental. p-values are given in parentheses (), and odds ratios (ORs) are given in curly brackets \{\}. ORs are calculated on the basis of one standard deviation change in the explanatory variable.}
\label{TAB:all}
\footnotesize{
\begin{tabular}{lcccc}
\hline
\multicolumn{ 5}{c}{\textbf{Panel A: Investor categories}} \\ \thickhline
           & {\bf Posts}       & {\bf Comments}  & {\bf Likes}  & {\bf Shares}             \\
\thickhline
Companies &      0.011         *** &   5.55E-06             &   7.61E-07          ** &  --6.40E-07             \\

    {\bf } & (3.71E-12)             &    (0.098)             & (6.44E-03)             &    (0.775)             \\

    {\bf } &    \{1.064\}             &    \{1.016\}             &    \{1.034\}             &    \{0.997\}             \\
\hline
Financial &   5.92E-03             &   1.25E-05             &   9.42E-07             &  --1.94E-06            \\

    {\bf } &    (0.185)             &    (0.197)             &    (0.226)             &    (0.758)             \\

    {\bf } &    \{1.034\}             &    \{1.036\}             &    \{1.042\}             &    \{0.991\}             \\
\hline
Governmental &      0.015             &   5.45E-05          ** &   2.62E-06             &   2.19E-05             \\

    {\bf } &    (0.091)             & (7.20E-03)             &    (0.112)             &    (0.087)             \\

    {\bf } &    \{1.086\}             &    \{1.165\}             &    \{1.121\}             &    \{1.104\}             \\
\hline
Nonprofit  &      0.033         *** &   3.76E-05          ** &   4.43E-06         *** &   3.03E-05          ** \\

    {\bf } & (2.27E-07)             & (4.79E-03)             & (2.23E-04)             & (1.84E-03)             \\

    {\bf } &    \{1.203\}             &    \{1.111\}             &    \{1.212\}             &    \{1.147\}             \\
\hline
Households &      0.011        *** &  --4.46E-06         *** &   1.92E-07          ** &  --9.04E-06         *** \\

           & (5.74E-83)             & (7.64E-07)             & (6.89E-03)             & (1.23E-41)             \\

           &    \{1.064\}            &    \{0.988\}             &    \{1.008\}             &    \{0.960\}             \\
\thickhline
                                                \multicolumn{ 5}{c}{{\bf Panel B: Activity groups of household investors}} \\
\hline

\thickhline
Active &   2.41E-04             &  --5.82E-06             &  --9.74E-07             &  --1.36E-05         ** \\

    {\bf } &    (0.942)             &    (0.386)             &    (0.065)             & (1.32E-03)             \\

    {\bf } &    \{1.001\}             &    \{0.984\}             &    \{0.959\}             &    \{0.940\}             \\
\hline
Moderate &   1.85E-03             &   4.11E-06             &   2.67E-07             &  --1.71E-06            \\

    {\bf } &    (0.083)             &    (0.071)             &    (0.140)             &    (0.225)             \\

    {\bf } &    \{1.011\}             &    \{1.012\}             &    \{1.012\}             &    \{0.992\}             \\
\hline
Passive &      0.012         *** &  --9.33E-06         *** &  --7.33E-07         *** &  --1.60E-05         *** \\

    {\bf } & (8.55E-55)             & (2.37E-10)             & (4.53E-10)             & (1.73E-49)             \\

    {\bf } &    \{1.072\}             &    \{0.974\}             &    \{0.969\}             &    \{0.930\}             \\
\hline
Inactive &      0.015         *** &   1.01E-05         *** &   1.84E-06         *** &   1.09E-07             \\

           & (5.12E-67)             & (5.19E-12)             & (1.24E-42)             &    (0.911)             \\

           &    \{1.088\}             &    \{1.029\}             &    \{1.083\}             &    \{1.000\}             \\
\hline
                                    \multicolumn{ 5}{l}{*** p $< 0.001$; ** p $< 0.01$; * p $< 0.05$} \\
\hline
\end{tabular} }
\end{table*}

\citet{grinblatt00} argue that, roughly speaking, finance and insurance institutions, as well as companies, can be viewed as the most sophisticated investor groups, as they generally take larger positions, have more resources to spend on research, and in many cases view investment as full-time career. In light of this, our findings indicate that more sophisticated investors are more independent of Facebook activities, as there is clearly no association between Facebook activities and decisions of financial institutions. Assuming that an investor's activeness is related to his/her sophistication, our findings on household activity groups supports the result that more sophisticated investors behave more independently of Facebook data. 

Facebook can be seen as a secondary information channel compared to first-hand official company announcements published on the exchange, and companies are likely to strategically select information disseminated in Facebook \citep{jung2017firms}. Nonprofit organizations and households, as arguably less sophisticated investors \citep{grinblatt00}, may allow their trading decision to be affected by Facebook posts and activity, especially if they have no access to professional data sources. In line with this view, \citet{ammann2017impact} find that the trading decisions of unsophisticated investors are affected by postings that do not contain value-relevant information on a social trading platform.

As our question is if the decisions of different investors are associated with the Facebook data, we are mostly interested in whether the regression estimates for the Facebook variables are statistically and economically significant, while the signs of the coefficients are not in the main focus.\footnote{The number of data points in the regression analysis is 332--341, which does not automatically lead to significant estimates as very large data samples do.} However, a couple of words about the signs of the estimates in Table \ref{TAB:all}. In Panel A, the signs for posts, comments, and likes are consistently positive, except comments for households. The signs for shares are both positive (governmental and nonprofit) and negative (companies, financial, households), though not all of them are statistically significant, which can explain the variation. Panel B with activity groups reports positive estimates for posts, but there is more variation for comments, likes, and shares as passive and inactive investors have negative estimates. Looking deeper into the reasons of these findings is out of the scope of this paper and left for the future research, as it would require semantic analysis.\footnote{Additionally, one could consider if observed associations can represent a reverse causality so that investors are not reacting to social media posts but companies are posting on Facebook in response to changes in investment behavior. However,  the reverse causality seems unlikely, because the information about numbers of traders changing their position is not public.}

\section{Summary and conclusion}\label{conc}

This paper gives the first empirical evidence that Facebook activities affect the trading of different investors differently. We provide evidence that the decisions of arguably less sophisticated investors---that is, households and nonprofit organizations---to increase or decrease shareholdings are clearly associated with Facebook data. At the same time, the decisions of financial institutions, which are likely to be among the most sophisticated investors in the market, are not associated with Facebook activity. Moreover, less active households' decisions are related to Facebook, while the decisions of more active ones are not, which gives additional evidence that the less sophisticated the investor, the more closely related the behavior is to Facebook. Given that Facebook is not a regulated information channel compared to first-hand official exchange releases, companies are likely to strategically select what information to disseminate in Facebook \citep{jung2017firms}. This suggests that less sophisticated investors, who may not have access to professional sources for financial data and news, may be driven by biased information. 

In the future research we are planning to do sentimental analysis on the posts and comments to give a more comprehensive picture of the reactions of different investors to Facebook activities. Concentrating only on Nokia may introduce some investor clientele bias, since the investors interested in Nokia may in general be more social media and technology savvy and follow the posts because of their inclination towards technology. At this point, we were only able to collect the data for Nokia, but in the future research we are planning to extend the sample to a wider variety of stocks. 

\section*{Acknowledgements}
We want to thank Hannu K\"arkk\"ainen and Jari Jussila for their valuable efforts to enable this project. The research leading to these results received funding from Tampere University of Technology through Networked Big Data Science and Engineering Center and TUT's doctoral school. This project also received funding from the European Union's Horizon 2020 research and innovation program under Marie Sklodowska-Curie grant agreement No. 675044. First author is grateful for the grants received from KAUTE Foundation and Nordea Bank Foundation sr. The funders had no role in study design, data collection and analysis, decision to publish, or preparation of the manuscript. 

\section*{References}
\footnotesize{
\bibliography{references}
}

\end{document}